\documentclass[]{aa}
\usepackage{graphics}
\begin{document}
\thesaurus{02.18.5;  08.19.5 SN 1006; 13.07.3; 13.25.5}
%\thesaurus{3 (11.02.2, 13.07.2)}
%
%
\title{On the origin of TeV radiation of SN 1006} 
\author{
F.A. Aharonian\inst{1},
A.M. Atoyan \inst{1,2},
}
\institute{Max Planck Institut f\"ur Kernphysik,
Postfach 103980, D-69029 Heidelberg, Germany \and
Yerevan Physics Institute, Alikhanian Br. 2, 375036 Yerevan, Armenia}

\offprints{F.A. Aharonian, email address: 
Felix.Aharonian@mpi-hd.mpg.de}
\date{Received 26 February 1999 / Accepted 9 September 1999}
\authorrunning{F.A. Aharonian and A.M. Atoyan}
\titlerunning {TeV emission of SN 1006}
\maketitle
\begin{abstract}
We discuss the link between the nonthermal X-radiation and TeV 
$\gamma$-ray emission from SN 1006, and  study the capabilities 
of both electronic and nucleonic models for  explanation of the TeV flux
observed from the northeast rim of SN 1006.      
We show that the interpretation of the TeV radiation by 
the inverse Compton scattering of electrons on 2.7 K  cosmic microwave  
background  radiation  is possible, however 
due to the escape of high energy electrons, 
the $\gamma$-ray emission should  be significantly contributed not 
only from the rim,  but  also  from the inner 
parts of the remnant. This  implies  an  angular size of the 
TeV $\gamma$-ray emission larger than the size of  the 
nonthermal X-radiation.  In this scenario the magnetic field in the rim 
should not exceed  $10 \, \mu \rm G$. Then,
 in order to allow acceleration of particles well beyond 10 TeV
the  shock speed should be high, $\geq 3000 \, \rm km/s$. 
The latter condition  gives preference to a large distance 
to the source, $d \simeq 2 \, \rm kpc$ or so which is 
in the limits  of distances currently discussed in the literature.  
On the other hand, a  larger magnetic field of order $100 \mu \rm G$ 
and  smaller shock speeds (and therefore a small distance 
to the source of about 1 kpc)  are not excluded.  In that case
the observed  TeV radiation can be explained 
by  shock  accelerated protons in the rim through 
production  and subsequent  decay of $\pi^0$-mesons. 
Contrary to IC radiation, the $\pi$-decay $\gamma$-ray source coincides 
essentially with the rim,  and therefore it could be recognized 
by a relatively compact angular width. Both the electronic and 
nucleonic models require high efficiency of particle acceleration close to the Bohm limit, and 
large total energy in accelerated particles at the level of respectively $1 \%$ 
and  $10 \%$ of the kinetic energy of explosion of SN 1006.  
We discuss observational possibilities  to distinguish  between 
electronic  and nucleonic origins of  $\gamma$-radiation.

\end{abstract}
\section{Introduction}

The flux of cosmic rays is described by a smooth 
single power-law spectrum which extends up to the so-called 
`knee'  around $10^{15} \, \rm eV$.  This can 
be interpreted as an evidence that the bulk of 
galactic cosmic rays is produced by a single source population. 
Since the early 60's the supernova remnants (SNRs) are believed to 
be the most probable sites of acceleration of galactic cosmic rays
(see e.g.  Ginzburg \& Syrovatskii \ \cite{GS64}),  the basic 
argument being that the galactic Supernovae (SNe), and 
the resulting supernova remnants (SNRs) are almost the only 
known potential sources with available (kinetic) energy needed  to provide 
the observed luminosity of the Galaxy in cosmic rays, 
$L_{\rm CR} \geq 10^{40} \, \rm erg/s$. Moreover, it has been shown  
(see e.g.  Drury  \cite{Drury83}, Blandford \& Eichler\ \cite{BlEich87}, 
Berezhko \& Krymski\ \cite{BerKrm88}, Jones \& Ellison\ \cite{JonEl91}) 
that a viable mechanism -- diffusive shock 
acceleration --  can effectively operate 
in relatively young SNRs accelerating particles to 100 TeV  
(Lagage \& Cesarsky \cite{LagCes83}) or perhaps to  higher energies
(V\"olk \& Biermann \cite{VlkBi}).   
The strong shocks in SNRs
provide not only very effective, 10 per cent or more,  conversion of
the total SN explosion energy  into
the accelerated particles   but also can naturally explain hard 
($\propto E^{-\Gamma}$ with $\Gamma \sim 2.0-2.1$)  source spectrum 
which follows from   the standard `leaky box' propagation 
model of  galactic  cosmic rays (see e.g. Gaisser \cite{Gaisser1}).

The nonthermal (synchrotron) radio emission observed from shell-type 
SNRs is an unambiguous indicator of acceleration  of GeV electrons there.
Possible association of some of the $\gamma$-ray 
`hot spots' detected by COS B  (Bhat et al. \ \cite{Bhat}; 
Pollock \cite{Pollock};  Wolfendale \& Zhang \cite{Wolf})
and EGRET (Sturner \& Dermer \cite{StrDer};  Esposito et al. \cite{Esposito}) 
with galactic SNRs could be another  evidence for acceleration 
of  electrons and  possibly also protons in SNRs  
(Blandford \& Cowie \cite{Bland}, Sturner et al. \cite{Sturner}; Gaisser et al.
\cite{Gaisser2}; 
Baring et al. \cite{baring}). However these observations concern only
the low energy, typically $E \leq 10 \, \rm GeV$ particles, therefore 
they cannot yet conclusively  argue in favor of SNRs as suppliers 
of the whole spectrum of galactic cosmic rays.  An unbiased proof 
of this hypothesis  can be provided only by  detection of  TeV 
$\gamma$-radiation produced at interactions of 
accelerated protons and nuclei with ambient gas through production 
and subsequent decay of $\pi^0$-mesons  
(Drury et al. \cite{DAV}; Naito \& Takahara \cite{NT}) 
and/or by accelerated   electrons upscattering  the 2.7 K MBR
(Mastichiadis \cite{masti0}).
Therefore it is difficult to overestimate the significance  
of the discovery  of TeV radiation from 
the historical supernova  SN 1006 reported by the CANGAROO 
collaboration  (Tanimori et al. \cite{tani}).

Actually the first, although rather  circumstantial evidence 
for the presence of ultrarelativistic particles in SN 1006  
was  found  two decades ago by means of X-ray observations. 
It was suggested that the power-law  X-ray spectrum  of SN 1006  
observed by the Einstein  
(Becker et al. \cite{Xeinst}) and EXOSAT (Jones \& Pye \cite{Xexosat}) 
satellites could be explained by synchrotron radiation of  
electrons accelerated in the  SNR shell (Reynolds \& Chevalier \cite{Chevalier}, 
Amosov et al. \cite{Amosov}). The recent detection
of a spatially resolved power-law component of hard X-radiation 
from the edges of the remnant of SN 1006
by ASCA (Koyama et al. \cite{Koyama}) and ROSAT (Willingale et al.  \cite{ROSAT})  
strongly supports the synchrotron origin of the radiation that implies 
acceleration of electrons to energies  $\sim 100 \, \rm TeV$.     
It should be noted that  a featureless, power-law spectrum of X-radiation 
could be produced by a plasma with 
multi-temperature structure (Hamilton et al. \cite{Hamilton}), or
by a  heavy-element-dominated plasma with possible deviation of
the electron distribution from the standard   Maxwellian 
spectrum (Laming \cite{Laming}).
Therefore the very fact of detection of a power-law component of X-radiation
does not yet  automatically  imply a nonthermal  origin of the radiation. However
in the case of SN 1006 the thermal interpretation of
the continuous X-radiation faces serious difficulties, and the
preference therefore is given to the nonthermal synchrotron hypothesis
(Koyama et al. \cite{Koyama};  Laming \cite{Laming}). 

Motivated by this fact,  several  theorists 
(Pohl \cite{pohl}; Mastichiadis \&
de Jager \cite{masti}; Yoshida \& Yanagita \cite{yoshida})  predicted 
detectable fluxes of  inverse Compton (IC) TeV emission.  
Remarkably, very soon this object has been detected  as a 
TeV emitter  (Tanimori et al. \cite{tani}). Both the  reported  
flux  and the  spatial position of the TeV 
source centered on the NE rim
are close  to the predictions. This circumstance 
strengthened the belief in the electronic   origin of TeV radiation.
The  $\pi^0$-decay  contribution  to the observed 
flux is widely considered as less important.

However,  given the fact that  SN~1006 is the only shell-type
SNR reported as TeV  $\gamma$-ray source, a  careful examination  
of different scenarios of $\gamma$-ray production 
is obviously needed prior to exclude the  nucleonic origin of the 
observed TeV radiation (Aharonian \cite{vertalk}).  
In particular, the contribution of the 
$\pi^0$-decay $\gamma$-rays could be enhanced to the level 
of the observed fluxes if we assume a relatively small distance 
to the source of about  $\simeq 1\, \rm kpc$
(Willingale et al. \cite{ROSAT}), as well as 
take  into account a significant compression of gas 
in the NE  rim by the shock. It is important that
this scenario does not limit the  magnetic field in the
acceleration region  to values comparable with the interstellar
B-field as required by the IC  models of the observed 
TeV radiation.  The magnetic field in the rim could be then as high as 
$B \sim 100 \mu \rm G$ which is favorable  for acceleration 
of particles up to $\gg  10 \, \rm  TeV$,  taking into account 
the young  age of SN~1006.  Obviously in both scenarios the nonthermal 
X-ray emission observed from the NE rim is 
explained by synchrotron radiation of directly accelerated electrons.  

The current data do not allow us to give preference to the electronic
or nucleonic origin of the observed TeV radiation.  
In  this paper we study in detail the spectral and spatial 
characteristics of the IC
and  $\pi^0$-decay $\gamma$-rays. 
In particular we show that while the size of $\pi^0$-decay 
$\gamma$-ray emission is  expected to be  
of the size of the  rim,  the size of the inverse Compton 
TeV source should be significantly larger because of   
inevitable escape of very high energy electrons from the   
acceleration region,   and their subsequent upscattering 
of the 2.7 K  MBR outside of the rim.

\section{Inverse Compton versus $\pi^0$-decay $\gamma$-rays} 

For a ``standard'' injection spectrum  of shock-accelerated particles 
(protons and electrons) in a SNR, 
\begin{equation}
Q(E)  \propto E^{-\Gamma} \, \exp{(-E/E_0)} \, ,
\label{standard}
\end{equation}
with $\Gamma \approx 2$ and $E_0 \approx 100 \, \rm  TeV$,
the flux of  $\pi^0$-decay $\gamma$-rays at 
$E \leq 0.05 E_0$,
\begin{equation}
F(> E) \approx 10^{-11} \, A \rm (E/1 \, 
{\rm TeV})^{-1}  \, \rm ph/cm^2 s \, ,
\label{intflux}
\end{equation}
is determined by a simple scaling parameter,
\begin{equation}
A=(W_{\rm CR}/10^{50} \, {\rm erg}) \, (n/1 \,
{\rm cm^{-3})} \, (d/1 \, {\rm kpc})^{-2} \, .
\label{A}
\end{equation} 
Here $W_{\rm CR}$ is the total energy in relativistic  protons, $n$ 
is the ambient gas density, and $d$ is the  distance to the source. 

The peak luminosity of $\gamma$-rays is reached   at 
the early Sedov phase (Berezhko \& V\"olk \cite{BVLK}), i.e.  
typically $10{^3}-10^{4}$ years after  the SN 
explosion. At this stage the  radius of the shell exceeds 
several parsecs. This results in  a typical angular size of
relatively close ($d \leq 1 \, \rm kpc$) SNRs of about $1^{\circ}$.
The apparent conflict, from the point of view of source 
detectability, between the angular size
($\phi \propto 1/d$) and the $\gamma$-ray flux 
($F_\gamma \propto 1/d^2$) significantly limits the 
number of SNRs   which could be detected by the current  
most sensitive imaging atmospheric Cherenkov telescopes (IACT), 
to the range of  $A \geq 0.1$. Future large stereoscopic IACT arrays 
will extend this limit down to $A \sim 0.01$ 
(Aharonian \& Akerlof \cite{AhAk}). 

%f1
\begin{figure}[htbp]
%\hspace*{7cm}
\resizebox{8cm}{!}{\includegraphics{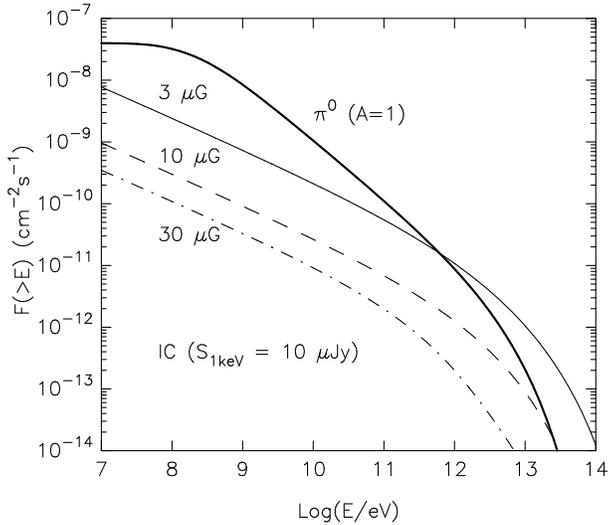}}
\caption
{The integral fluxes of $\pi^0$-decay (heavy solid line) 
and IC  (thin lines) $\gamma$-rays
from a $10^3 \, \rm year$ old SNR. The $\pi^0$ decay
$\gamma$-rays are calculated for the scaling parameter $A=1$.
The IC $\gamma$-ray fluxes are calculated for 3 different values  of
the magnetic field $B=3$ (solid curve), 10 (dashed curve),  and
$30 \, \rm \mu G$ (dot-dashed curve),  assuming the electrons
produce the same flux of synchrotron radiation $S_\nu=10 \, \mu \rm Jy$. 
For both the protons and electrons we assume the same injection
spectrum given in the form of Eq.(1) with $\Gamma=2$ and
$E_0=100 \, \rm TeV$.}
\label{f1}
\end{figure}

When extracting information about the accelerated protons one has to 
subtract  a possible non-negligible ``contamination'' caused  by  
parallelly accelerated electrons that upscatter the photons of 
2.7 K  MBR (which is the dominant target photon field for production 
of TeV photons in most of SNRs; see e.g. Gaisser et al. \cite{Gaisser2}) 
up to $\gamma$-ray energies.
For production of IC 
$\gamma$-rays of TeV energies multi-TeV  electrons are needed. 
The same electrons produce also synchrotron UV/X-ray 
radiation. The typical energies  $E_\gamma$ and $E_{\rm X }$ 
of the IC and synchrotron photons produced by an electron are 
related as (see e.g. Aharonian et al. \cite{AAK}) 
\footnote{Note that this  relation neglects the 
Klein-Nishina effect which becomes important at energies 
$E_\gamma  \geq 10 \, \rm TeV$.} 
\begin{equation}
E_\gamma \simeq 1.5 (E_{\rm X}/0.1 \, \rm keV) \, (B/10 \, 
\mu \rm G)^{-1}  \, \rm TeV \, .
\end{equation}
The ratio of the synchrotron and IC fluxes $f_{\rm E} \equiv E^2 F(E)=\nu S_\nu$ 
at these energies does not practically depend on the shape of the spectrum
of parent relativistic electrons, but strongly depends on
the magnetic field:
\begin{equation} 
\frac{f_{\rm IC} (E_\gamma)}{f_{\rm sy} (E_{\rm X})} 
\simeq 0.1 \,  (B/10 \, \mu \rm G)^{-2}
\end{equation}

For the X-ray spectrum with photon index 
$\alpha_{\rm X} \sim 2$ the energy flux $f_{\rm X}$ 
is almost energy-independent. Therefore 
$f_{\rm X}$  at  a  typical  energy  of 1 keV
could serve as a good indicator for the IC $\gamma$-ray 
fluxes expected at TeV
energies, although for magnetic fields $B \leq 100 \, \mu \rm G$ 
the energy of synchrotron photons relevant 
(produced by the same parent electrons) 
to $\sim 1 \, \rm TeV$ 
$\gamma$-rays is in the soft X-ray domain (see Eq. 4).
    
The contribution of $\pi^0$-decay $\gamma$-rays dominates over 
the  contribution of the  IC  component  when 
\begin{equation}
A \geq  0.1 \, (S_{1 \rm keV}/{10 \mu \rm Jy}) \, 
(B/10 \, \mu \rm G)^{-2}~,
\end{equation} 
where $S_{1 \rm keV}$ is the flux of nonthermal synchrotron radiation 
at 1 keV normalized to  $10 \, \mu \rm Jy$ (the corresponding energy flux 
$f_{\rm X} \approx  2.4 \times 10^{-11} \, \rm erg/cm^2 s$). Note that   
$S_{\rm 1keV}=10 \, \mu \rm Jy$ is a typical level of nonthermal X-ray fluxes 
reported recently from four  shell-type SNRs --
SN 1006 (Koyama et al. \cite{Koyama}),  Cas A (Allen et al. \cite{Cas}), 
IC 443  (Keohane et al. \cite {IC443}) and  RXJ1713.7-3946 
(Koyama et al. \cite{Koyama2}).

In Fig.~\ref{f1} we present integral fluxes of  
$\pi^0$-decay and inverse Compton $\gamma$-rays from a  SNR
of age $10^{3} \, \rm yr$. 
The $\pi^0$-decay $\gamma$-ray flux
corresponds to the scaling factor $A=1$. The IC fluxes are calculated by  
normalizing the  synchrotron X-ray fluxes  to $S_{\rm 1 keV}=10 \mu \rm Jy$
for ambient B-fields $3 \mu \rm G$, $10 \mu \rm G$, and $30 \mu \rm G$. For
both electrons and protons we assume continuous acceleration during
$10^{3} \, \rm years$ with a time-independent source spectrum taken in  
the form of Eq.(1) with $\Gamma=2$ and
$E_0=100 \, \rm TeV$.  Note that for the normalizations used,
the results presented in Fig.~\ref{f1}   only slightly depend on the source age, 
unless it is larger than the radiative cooling time of multi-TeV electrons, 
$t_ {\rm rad} \approx 10^4 \, (B/10 \, \mu \rm G)^{-2} \, 
(E_{\rm e}/10 \, \rm TeV)^{-1} \, \rm yr$.

\section{Synchrotron and IC components  of 
nonthermal radiation of SN 1006}     

The   bulk ($\simeq 75 \%$) of the nonthermal X-ray component  of SN 1006
is contributed by the northeast (NE)  and the southwest (SW) rims 
(Koyama et al. \cite{Koyama},  Willingale  et al. \cite{ROSAT}).  
The observations of SN 1006 by the CANGAROO
collaboration (Tanimori et al. \cite{tani}) have revealed TeV radiation
from the  NE rim, but only an upper limit from the SW rim was reported.
Therefore in this paper the nonthermal radiation of only NE rim is
discussed.

For the spectral characteristics of the nonthermal X-ray emission of
NE rim we follow the procedure used by Mastichiadis \& de Jager  (\cite{masti}).
Namely we assume that the X-ray spectrum below 1.5 keV is described by a  power-law
with mean photon index $\alpha_{\rm x} \simeq 2.2$ as it follows from the
combined Einstein/EXOSAT/ROSAT data. At higher energies up to
8 keV the spectrum reported by ASCA from NE+SW rims is significantly 
steeper,  with a power-law photon index $\alpha_{\rm x} \simeq 3$. 
For the estimate of the absolute X-ray flux from NE rim we
take into account that  $\simeq 60 \%$ of the 
total emission of NE and SW rims  is contributed by NE rim,
as it  follows from the respective count rates of ROSAT 
(see Willingale et al. \cite{ROSAT}).
The resulting range of the X-ray fluxes observed from the NE rim 
in the range from 0.1 keV to 8 keV  is shown in Fig.~\ref{f2}.

The radio emission of SN 1006   has a power-law spectrum with
an index $\alpha_{\rm r} \simeq 0.57$ and intensity 30.8 Jy at 408 MHz 
(Stephenson  et al. \cite{radio}). In Fig.~\ref{f2} one half  of this flux is 
taken because the observed total radio flux is equally
contributed from the northeast and southwest directions 
(see e.g. Reynolds \& Gilmore \cite{ReGil86}). 

In Fig.~\ref{f2}  we show also the range of TeV $\gamma$-ray fluxes observed
by CANGAROO from the direction of  NE rim of SN 1006 at energies above 1.7 TeV 
(Tanimori et al. \cite{tani}), as well as  the EGRET  flux upper 
limit   from  SN 1006 (Mori \cite{Mori}).  

\subsection{One zone model}

In this section we assume that 
synchrotron and IC $\gamma$-rays are produced
by relativistic electrons confined 
in a spatially homogeneous rim. In this simplified
case the IC fluxes shown in Fig~1 could serve as  first 
approximation estimates for the expected TeV $\gamma$-radiation.
However, more reliable predictions of $\gamma$-ray 
fluxes require a careful derivation  of the spectrum of  
$\geq 10 \, \rm TeV$ electrons which is controlled by
the spectrum of synchrotron X-rays.

%f2
\begin{figure}[htbp]
%\hspace*{7.cm}
\resizebox{8.cm}{!}{\includegraphics{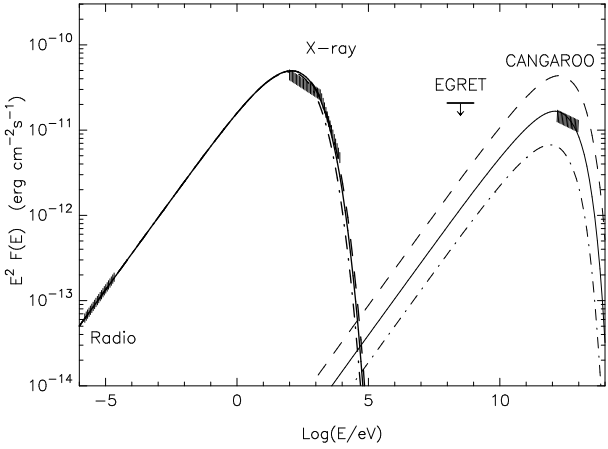}}
\caption
{The synchrotron (heavy lines) and IC (thin lines) 
fluxes calculated for the  homogeneous source without escape of 
accelerated particles (one-zone model).  Power-law  
injection spectrum of electrons with $\Gamma_{\rm e}=2.15$
is assumed in order to fit the radio data. 
The maximum energy $E_0$ is determined from the condition 
given by Eq.(9) for 3 different values of the magnetic field:
$B=3 \, \rm \mu G$ (dashed line), $5 \, \rm \mu G$  
(solid line), and $8 \, \rm \mu G$ (dot-dashed line).}  
\label{f2}
\end{figure}

In the synchrotron-inverse Compton models
the spectral fit to the  X-ray flux is crucial
because the fluxes  are produced  by electrons
in the region of the exponential cutoff $E_0$  
between 10 and 100 TeV. 
The spectral index $\Gamma$ of accelerated electrons 
is derived  from the radio data, 
$\Gamma=1+2 \, \alpha_{\rm r} \simeq 2.15$. Information about $E_0$ 
in Eq.~(1) is contained in the  X-ray spectrum .
In particular case of negligible energy losses, the spectrum of
electrons  $N(E_{\rm e})$ repeats the injection spectrum, 
$N(E_{\rm e}) \propto Q(E_{\rm e})$,
therefore in the $\delta$-functional approximation for 
the synchrotron emissivity, the X-ray spectrum can be 
presented in a  simple form (see e.g. 
Yoshida \& Yanagita \cite{yoshida}, Reynolds \cite{Reyn})
\begin{equation}
F(E) \propto E^{-(\Gamma +1)/2} \, \exp[-(E/E_{\rm m})^{1/2}] \, , 
\end{equation}
where  
\begin{equation}
E_{\rm m} \simeq 0.29 h \nu_{\rm c} \simeq 0.02 \, (B/10 \mu {\rm G}) \, 
(E_0/10 \, \rm TeV)^2 \, \rm keV 
\label{emax}
\end{equation}
is the mean energy of synchrotron photons
produced by an electron of energy $E_0$ 
($\nu_{\rm c}$ is the characteristic synchrotron frequency ).

From Eq.~\ref{emax} follows that the spectrum of synchrotron radiation 
depends only on the product $E_0 \, B^{1/2}$.  
Numerical calculations using the accurate expression for synchrotron 
radiation  generally confirm that approximation in the form 
of Eq.(7) holds at least until $E \sim10  E_{\rm m}$ (Reynolds \cite{Reyn}).

In Fig.~2. we show 3 spectra of synchrotron radiation
calculated for the same product 
\begin{equation}
\Pi=E \, B^{1/2}=61.5 \, \rm TeV \, \mu G^{1/2} \, ,
\end{equation}
but for 3 different combinations of $E_0$ and $B$:
(1)  $B=3 \mu \rm G$, $E_0=35.5 \, \rm TeV$;
(2)  $B=5 \mu \rm G$, $E_0=27.5 \, \rm TeV$;
(3)  $B=8 \mu \rm G$, $E_0=21.3 \, \rm TeV$ 
\footnote{
Note that our best fit value for the product 
$\Pi=E \, B^{1/2}=61.5 \rm TeV \, \mu G^{1/2}$ 
is by a factor of two less than 
obtained by Yoshida \& Yanagita \cite{yoshida}. The reason presumably is 
connected with different fluxes of X-rays used in calculations.}. 
In all calculations  we use the same normalization to
the radio fluxes. Therefore different
magnetic fields require different total energy
in relativistic electrons.  
At the same time, because the target photon field (2.7 K MBR) 
for the IC scattering is fixed, there is a strong dependence
of the IC $\gamma$-ray fluxes on the magnetic field. In particular,
in the case of small magnetic fields $B \leq 10 {\mu \rm G}$, 
when the radiative cooling time 
\begin{equation}
t_{\rm sy}\approx 1.2 \cdot 10^{4} 
\, (E/10 \, {\rm TeV})^{-1} (B/10 \, {\mu \rm G})^{-2} \, \rm yr    
\end{equation}
of 10 to  100 TeV electrons  responsible for the 
observed X-rays and TeV $\gamma$-rays   
exceeds the age  of the source,  we have
$ F_\gamma \propto B^{-(\alpha_{\rm r}+1)}$.
Note that for $B \leq 3  {\mu \rm G}$ the radiative losses 
on 2.7 K MBR become dominant.   

Fig.~\ref{f2} shows that the interpretation of the observed TeV radiation
by IC mechanism in the framework of a simplified  spatially  
homogeneous (one-zone) model is possible only for an 
ambient magnetic field in a very narrow range 
around  $5 \mu \rm G$.  Magnetic fields $\geq 7 \mu \rm G$
result in a strong  reduction of the IC flux, 
while the assumption of low magnetic fields, $B \leq 4 \, \mu \rm G$,
leads to overproduction of $\gamma$-rays.

For such small magnetic fields a question arises whether the 
acceleration of electrons could be efficient enough in order
to boost  particles up to energies 
$E_0=\Pi /B^{1/2} \sim 25-30 \, \rm TeV$
during the time of acceleration $t_0$ limited 
by the  age $10^3$ yr of SN 1006.    
Indeed, in the model of diffusive shock acceleration the 
characteristic maximum energy $E_0$  
that a particle could achieve during the time $t_0$
is determined (see e.g. Lagage and Cesarsky \cite{ LagCes83}) 
by the shock velocity 
$v_{\rm s}$, the magnetic field $B$, and the so-called gyrofactor 
$\eta$ (the ratio of the particle mean free path to the gyroradius):
\begin{eqnarray}
E_{\rm 0} & \simeq & 20 \, \left( \frac{B}{10 \, \rm \mu G} \right) \, \left( 
\frac{t_0}{10^3 \, \rm yr} \right) \times  \nonumber \\ 
& & \left( \frac{v_{\rm s}}{2000 \, \rm km/s} \right)^2 \, \eta^{-1} \, \rm TeV \, . 
\end{eqnarray}
This estimate would be not  significantly  changed even
if we take into account that in the past the shock speed was much
higher than the present value $v_{\rm s}$ in Eq.(11). Indeed, even
assuming that the Sedov phase in SN 1006 has started at $t_\ast \ll 1000 \, \rm yr$
and applying the Sedov solution for the shock speed in the past, 
$v(t)=v_{\rm s} (t/t_0)^{-3/5}$ for $t \geq t_\ast$, the maximum energy
of accelerated particles which is reached at $t \sim t_\ast$ could be
higher than it  is given by Eq.(11) only  by a factor of  $(t_0/t_\ast)^{1/5}$. 
Therefore, even for $t_\ast \sim 0.1 t_0 \sim 100 \, \rm yr$, the maximum energy
of the particles in the past  could be only by a factor of 1.5 higher
than it follows from Eq.(11).  On the other hand  Eq.(11) does not include 
possible dissipative effects, such as  the wave damping and the  synchrotron 
losses of  the electrons,  therefore it can be considered as a good  estimate for 
the maximum possible energy of accelerated particles.  

Eqs.(9) and (11) result in a lower limit on the
strength of the magnetic field
\begin{equation}
B \geq 12 \,  \left( \frac{t_0}{10^3 \, \rm yr} \right)^{-2/3} \,
\left( \frac{v_{\rm s}}{2000 \, \rm km/s} \right)^{-4/3} \, 
\eta^{2/3} \, \rm \mu G
\end{equation}

The  Sedov solution for the shock speed in SN 1006 gives 
(Willingale et al. \cite{ROSAT}) 
\begin{equation}
v_{\rm s} \simeq 2000 (d/1 \, {\rm kpc})^2 \, \rm km/s . 
\end{equation}
 
The shock speed  $1400 \pm 200 \, \rm km/c$ 
estimated  by Willingale et al. (\cite{ROSAT}) from the ROSAT
X-ray data implies a rather small distance to the source,
$d=0.7 \pm 0.1 \, \rm kpc$. 
Winkler \&  Long  (\cite{winkler}) argued that the 
distance to the source should be larger,  $d=1.8 \pm 0.3 \, \rm kpc$. 
At the same time, the shock speed  $v_{\rm s}=2600 \, \pm 300 \rm km/c$
preferred by Winkler \&  Long  (\cite{winkler}) is significantly lower than
$\simeq 3600\,\rm  km/c$ following from Eq.(12) at $d=1.8\,\rm kpc$. 
Given the current controversy both in the estimated shock speed and the distance,
below we will discuss distances in the range from 0.7 to 1.8 kpc, but using 
for all distances  Eq.(13).  

Thus,  if the shock speed does not significantly exceed 2000 km/s 
which implies a distance $\simeq 1 \, \rm kpc$, even in the 
Bohm limit ($\eta=1$) the strength of the  magnetic field should 
not  be less than $10 \, \rm \mu G$ even if we assume that 
the effective particle acceleration has started at early stages of this SNR.
This limitation could be softened by a factor of 2  for larger 
distances to the source  (see below).

Another way to soften the 
conflict between the sestimate of the B-field 
given by Eq.(12) and the value of $B \leq 7\, \rm \mu G$ needed for
explanation of TeV fluxes  is possible if we assume   
that the bulk of the observed $\gamma$-ray flux is  produced outside of the shell. 
The  highest energy electrons produced in the shell 
can effectively escape  the  acceleration region. This will give rise to an 
enhanced IC  emission outside of the rim, namely in the interior regions of the
remnant  where the 
magnetic field could be as low as the typical interstellar B-field. 
Such a leakage of electrons from the rim is unavoidable
because of diffusive and convective  propagation of particles.  
This effect should be taken into account  
not only for correct derivation of the electron spectrum inside the rim,  
but also for calculations of the IC  radiation 
outside of the rim.  Below we show that in fact the contribution of this
component (missed in previous studies)  is comparable or even can
exceed  TeV flux produced in the rim.

\subsection{Inverse Compton  TeV emission outside of the rim} 

The radio and X-ray images of SN 1006 
indicate that the most probable  site of particle acceleration 
is the  thin shell of the remnant, in particular the NE rim.
The width $\Delta r$ of the NE rim in X-rays  as measured by ROSAT
does not exceed $20 \%$ of the angular radius of the remnant
of about 17 arcminutes  (Willingale et al. \cite{ROSAT})
which corresponds to
$r_{\rm s}=4.9 \, d_{\rm kpc} \, \rm pc$  at a distance to the source 
$d=1 \, d_{\rm kpc} \, \rm kpc$. 
In calculations we assume a constant  $\Delta r/r_{\rm s}=0.2$ 
ratio throughout the  evolution of the remnant. 

Due to two principal mechanisms of propagation -- diffusion and convection --
the relativistic particles cannot be completely confined in the rim.
The characteristic escape time  is $\tau_{\rm esc}=(\tau_{\rm con}^{-1}+
\tau_{\rm dif}^{-1})^{-1}$. The convective escape time is 
\begin{equation}
\tau_{\rm con}(t)=\Delta  r/u_2 \, ,
\end{equation}
where $u_2$ is the fluid speed 
downstream of the shock (in it's rest frame), thus
at present $ u_2=v_{\rm s}/\rho \sim 500 \, \rm km/s$ 
for the compression ratio $\rho \sim 4$.
The  width of the shell $\Delta r \simeq 0.2 r_{\rm s} \propto t^{2/5}$ 
assuming a  Sedov-type solution for the shock radius $r_{\rm s}$.   
Note that strictly speaking this solution is valid only at stages after
the onset of the Sedov phase,  i.e.  for $t_\ast \le  t \leq t_0$.
However the energy distribution of electrons formed in the shell 
at present should be  dominated by particles accelerated
recently, $t \geq 0.3 t_0$ or so, unless one would assume 
that most of particles have been injected at very early stages.
Therefore possible deviation
of $\Delta r(t)$ from the law $\propto t^{2/5}$ at times
$t \ll t_0$  does not significantly affect the spectrum of electrons at
present.   

The diffusive escape time is
\begin{equation}
\tau_{\rm diff} (t) \simeq \frac{(\Delta r)^2}{2 \, D(E)} \, ,
\end{equation}
where $D(E)$ is the diffusion coefficient in the shocked region (the rim).
In the theory of shock acceleration 
the diffusion coefficient is generally taken in the form
\begin{equation}
D(E)=\eta \frac{r_{\rm g}\, c}{3} \, , 
\end{equation}
where $r_{\rm g} = 3.3 \cdot 10^{15}   
(E/10 \, \rm TeV)  (B/10 \, \mu \rm G)^{-1} \, \rm cm$ 
is the particle gyroradius, and $\eta \geq 1$ is the 
gyrofactor. The maximum confinement (and therefore the maximum energy) 
of  particles is achieved in the Bohm diffusion
regime corresponding to  $\eta=1$. 

The kinetic equation for energy distribution
of relativistic electrons $N(E)$ in a source reads 
(see e.g. Ginzburg \& Syrovatskii \cite{GS64}):
\begin{equation}
\frac{\partial N}{\partial t} = \frac{\partial}{\partial E_{\rm e}}
 \left[ P\, N \right] - \frac{N}{\tau_{\rm esc}} + Q \, ,
\end{equation}
where $Q\equiv Q(E_{\rm e}, t)$ is the electron injection rate, 
and   $P = -{\rm d} E_{\rm e}/{\rm d}t$  
is the electron energy loss rate.  
The solution to this equation, including the {\em time} and {\it energy}
dependent escape losses is  
(Atoyan \& Aharonian \cite{aa99}):
 
\begin{eqnarray}
N(E_{\rm e},t) & = & \frac{1}{P(E_{\rm e})} \int_{-\infty}^{t}
P(\zeta_t) Q(\zeta_t,t_1) \times \nonumber \\
& & \exp \left( -\int_{t_1}^{t}\frac{{\rm d} x}
{\tau_{\rm esc}(\zeta_x,x)}\right)
{\rm d} t_1\, .
\end{eqnarray}
The  variable $\zeta_t$ corresponds to the initial energy of an electron at 
instant $t_1$ which is cooled down to given energy $E_{\rm e}$ by the 
instant $t$, and is determined from the equation
\begin{equation}
t-t_1 =\int_{E_{\rm e}}^{\zeta_t}{\frac{{\rm d}E}{P(E)}} \, .
\end{equation}

In this paper we 
consider only  overall  (i.e. integrated over the volumes)  
fluxes from the NE rim (zone 1) and outside (zone 2).
The electron distribution in the rim $N_1(E_{\rm e},t)$ is found
assuming injection of electrons
at a constant rate during the age of SN~1006,  with the spectrum 
 $Q_1(E_{\rm e})$ in the form of Eq.(1).  Although  the injection rate of 
accelerated electrons is actually time-dependent,  the  
main contribution to the total 
amount of particles  in the rim comes from the acceleration
at  epochs from $t \sim t_\ast$ to 
several times $t_\ast$ ($t_\ast$ corresponding  to the  
onset of the Sedov phase) when the injection rate $Q_1(t)$ is almost
time-independent   (see Berezhko \& V\"olk \cite{BVLK}). 
Since SN 1006  most probably has already reached the Sedov phase
(V\"olk \cite{KPVLK})  an assumption of  time-independent  
$Q_1(E,t)$ is an adequate approximation (see below).  At the same time  
we assume time-dependent escape, as described above (see Eq.~14 and 15),
which have a stronger impact on the results
than the weak time dependence of the injection rate. 

Outside the rim the energy distribution 
$N_2(E_{\rm e},t)$ is found
from the same Eq.(18) where the injection function is given
by the rate of escape of electrons from the rim,
$Q_2(E_{\rm e},t)=N_1/\tau_{\rm esc}$. 
The region outside of the rim corresponds to the   
interior of  the remnant close to NE rim.  Therefore  in this region 
we ignore the losses due to escape.

%f3
\begin{figure}[htbp]
%\hspace*{7cm}
\resizebox{8.cm}{!}{\includegraphics{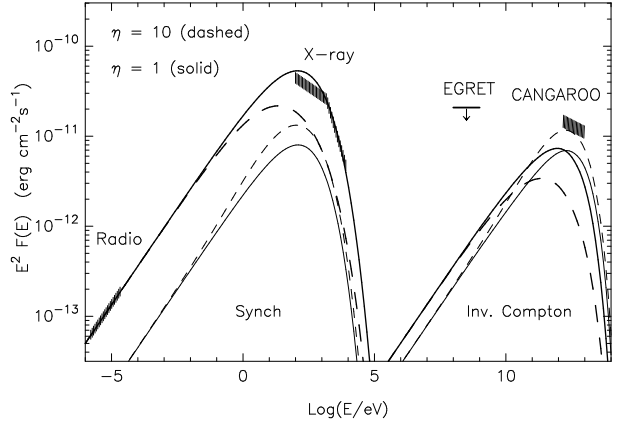}}
\caption
{The synchrotron  and IC  fluxes produced
inside (heavy lines) and outside (thin lines) of the rim 
calculated within 2-zone model for two values of the gyrofactor:
$\eta=1$ (solid), and 10 (dashed).  
The magnetic fields inside and outside 
of the rim are taken   as $B_1=8 \, \rm \mu G$ 
and   $B_2=3 \, \rm \mu G$, respectively. The distance to the source is assumed
1 kpc.}
\label{f3}
\end{figure}

In Fig.~\ref{f3}  we show the synchrotron and IC fluxes from the NE rim 
and from the inner part of the remnant  for the magnetic field in the rim $B_1=8  \, \mu  
\rm G$ and $B_2=3  \, \mu \rm G$ in the remnant. In order to show the 
impact of diffusive escape on the resulting radiation spectra, we
considered 2 values for the parameter $\eta$. 
The case of maximum confinement of particles in the rim,
$\eta=1$,  is shown  by solid lines, while the dashed lines correspond
to the case of $\eta=10$. The heavy and thin lines 
correspond to the  emission produced inside and outside of the rim, 
respectively.   
 
For the parameters described above 
the observed X-ray spectrum of the rim is best explained for
the exponential cutoff energy  $E_0 = 30 \, \rm TeV$. 
The increase  in the value of $E_0$ compared with the best-fit value
of 21.3 TeV used in Fig.~2 for the same magnetic field 8 $\mu \, \rm G$
in the simple single-zone approach, is    
explained by diffusive escape of the electrons from the rim. 
For the magnetic
field used,  even in the Bohm limit  the diffusive escape time of 
electrons  at present is $\tau_{\rm dif}\simeq 3500 \,\rm yr /(E/10 \,\rm TeV)$,
therefore for electrons with energy $E\sim (30-100) \,\rm TeV$
which are responsible for the X-ray fluxes, the diffusive escape time 
becomes comparable or even less than the age of the source. 
This results in a  modification (steepening) of the spectral 
shape at high energies.  In order to compensate this effect,
a higher $E_0$ is to be assumed. Note that 
the convective escape time at present,
$\tau_{\rm con} =2000 \,\rm yr$,
is also comparable with the 
age of the source.  However the convective escape does not 
modify the energy spectrum of electrons.
This is true, to some extent,  also for radiative losses in the case
of small magnetic fields   $B\sim 10\,\rm \mu G$,  
since the synchrotron cooling time then equals the
source age only at $E_{\rm e}\geq 100\,\rm TeV$. 

The electron escape from the rim leads to nonthermal X-ray
production in  a broader region (zone 2). Note that for the case 
of diffusion in Bohm limit ($\eta=1$) and the assumed magnetic field
$B_2=3 \, \rm \mu G$, the 
X-ray flux outside of the NE rim makes 
$\simeq 10 \%$ of the flux produced in the rim.
This implies that the magnetic field $B_2$ cannot be larger than 
$5 \, \rm \mu G$, otherwise the ratio of the fluxes 
produced outside and inside of the rim would
exceed the value of about 1/3 as observed by ASCA (Koyama et al. \cite{Koyama})
and ROSAT  (Willingale et al. \cite{ROSAT}).  

While the synchrotron radiation outside of the rim can be suppressed 
assuming low magnetic field $B_2$,  the formation of an extended 
IC radiation is unavoidable due to the uniform distribution of
the principal target photon field for the Compton scattering - 
2.7 K  MBR.  In particular even for $\eta=1$ 
the contribution of $\gamma$-ray fluxes at $\geq 1 \, \rm TeV$
from the region outside of the rim becomes comparable with the flux from 
the rim.  In the case of fast diffusion corresponding to  $\eta=10$,
the contribution from the extended region 
outside  of the rim,  i.e. inside the remnant at $r \leq 0.8  r_{\rm s}$,  
well dominates the overall TeV flux of SN~1006 (see Fig.~\ref{f3}).
   
The value of $E_0=30 \, \rm TeV$ in Fig.~3  
is by a factor 1.5 larger than the estimate of the 
characteristic maximum energy of electrons
following from Eq.~(11) for the shock speed 
2000 km/s corresponding to a distance to the source  
$d=1 \, \rm kpc$  (see Eq.~13).  
This conflict could be overcome assuming a higher speed of the shock, i.e.
lager $d$. It is important to note that the assumed distance to the source
has a significant impact also on the timescale of the diffusive 
(but not the convective)  escape of the electrons from  the shell as 
$\tau_{\rm dif} \propto \Delta r^2 \propto d^2$ (see Eq.~15).
Therefore the relative contribution of the TeV emission produced in the
interior of the remnant depends on the assumed distance to the source.
This effect is seen in Fig.~4 where we show the results of calculations
for two different distances currently discussed in the literature, 
$d=0.7 \, \rm kpc$ (Willingale et al. \cite{ROSAT}) and 
$d=1.8 \, \rm kpc$ (Winkler \& Long  \cite{winkler}).
For  the distance $d=0.7 \, \rm kpc$ we used 
$E_0=18 \, \rm TeV$ which is 2 times larger than it follows from
the estimate of the maximum energy by  Eq.~(11).  
However, even under such an extreme assumption the calculated synchrotron fluxes
roll over  too early, failing to reach the observed X-ray fluxes.
Meanwhile for the distance $d=1.8 \rm kpc$ the chosen value
of $E_0=35 \, \rm TeV$ is in agreement with Eq~(11) even for $\eta=2$.
We note that the contribution of the TeV flux produced outside of the rim 
is still very significant even for this  large distance.   

%f4
\begin{figure}[htbp]
%\hspace*{7cm}
\resizebox{8.cm}{!}{\includegraphics{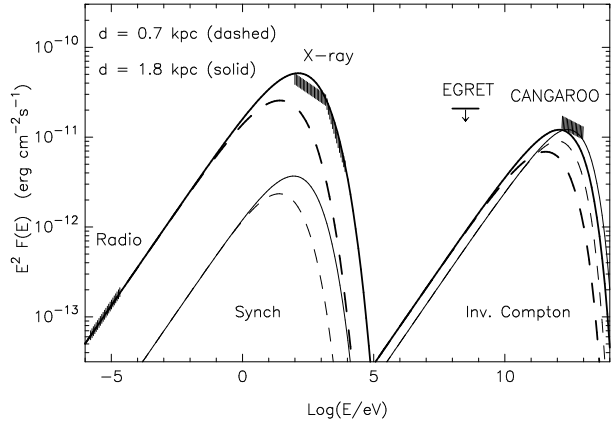}}
\caption
{The synchrotron  and IC  fluxes produced
inside (heavy lines) and outside (thin lines) of the rim 
calculated within 2-zone model for two distances to the source:
$d=1.8 \, \rm kpc$, and  $d=0.7 \, \rm kpc$.
The gyrofactor and maximum energy are: 
$\eta=2$ and $E_0=35 \, \rm TeV$  for $d=1.8 \, \rm kpc$
(solid lines), 
and $\eta=1$ and $E_0=18 \, \rm TeV$  for $d=0.7 \, \rm kpc$
(dashed  lines).   
The magnetic fields inside and outside 
of the rim are taken   as $B_1=6 \, \rm \mu G$ 
and   $B_2=1.5 \, \rm \mu G$, respectively. }
\label{f4}
\end{figure}

The existence of the external  IC  component due to the 
escape of electrons from the rim results in a significant increase 
of the overall TeV radiation, and thus allows a larger  magnetic 
field as compared with  the  value $B_1 = 5 \pm 1 \, \rm \mu G$ derived 
within the one-zone model (Section 3.1). In order to investigate
the impact of this effect we have calculated the synchrotron and IC 
fluxes   for two different combinations of $B_1$  and $\eta$ 
shown in  In Fig.~5. In both cases we assume for the distance to the source
$d=1.8 \, \rm kpc$.  The maximum electron energy 
$E_0=32 \, \rm TeV$ is calculated from Eq.(11).  
We also assume low  magnetic field 
outside of the rim,  $B_2=B_1/4$,  in order to avoid overproduction of 
X-radiation in this region.  

%f5
\begin{figure}[htbp]
%\hspace*{7cm}
\resizebox{8.cm}{!}{\includegraphics{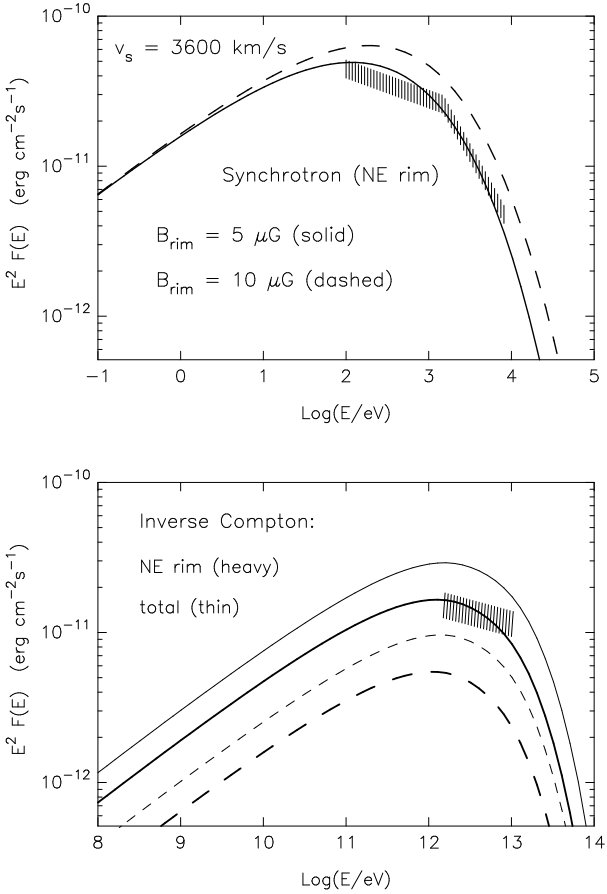}}
\caption
{The synchrotron (top panel) and IC  $\gamma$-ray (bottom panel)
fluxes calculated in the framework of 2-zone model for two values
of the magnetic field in the rim: $B_1=5 \, \rm \mu G$ (solid lines) and
$B_1=10 \, \rm \mu G$ (dashed lines). The magnetic field outside of the rim
is taken as $B_2 =B_1/4$. The heavy lines show the
fluxes from  the NE rim, and thin lines  correspond to
the total fluxes. The distance to the source is assumed 1.8 kpc. 
The maximum electron energy $E_0$ is calculated from Eq. (11)
for the gyrofactor $\eta=1$ for $B_1=5 \, \rm \mu G$, and
$\eta=2$ for $B_1=10 \, \rm \mu G$.}  
\label{f5}
\end{figure} 

The combination $B_{\rm rim} \equiv B_1=5 \, \mu \rm G$,
$\eta=1$ provides a good fit to the X-ray data,  but it 
leads to an overproduction of the total TeV 
emission compared with the CANGAROO measurements by a factor of
two. Given a possibility that the TeV emission from extended region 
of the remnant (outside of the rim), which contributes half of the total
flux,  could be difficult to extract   by the CANGAROO,  this combination 
of parameters seems still  acceptable.  The assumption of 
$B_1=10 \, \rm \mu G$ and $\eta=2$ results in overproduction 
of X-rays by $50 \%$. At the same time the predicted fluxes of TeV 
$\gamma$-rays from the rim are below the reported fluxes by
a factor of 3.    

From Fig.~5 we may draw a  conclusion  that the magnetic field 
in the rim should be within the limits $5 \le B_1 \le 10  \, \rm \mu G$
(preferably   $B_1 \simeq 6-8  \, \rm \mu G$), and $\eta$ close to 
1,  e.g. the diffusion in the rim should take place essentially  in the Bohm limit.
In order to provide maximum electron energy $E_0$ of order of 30 TeV, the
shock speed  should exceed 3000 km/s which implies a large 
distance to the source,  $d \geq 1.5 \, \rm kpc$. For this set of 
parameters, approximately half of the total TeV emission is contributed 
from the inner parts of the remnant,  
$r \leq 0.8 r_{\rm s}$. The narrow range of the required
magnetic field in the rim allows rather accurate estimate of the total energy
of relativistic electrons in SN 1006. For example for the distance
$d=1.8 \, \rm kpc$ in Fig.~4 the total electron energy in the rim is 
$W_{\rm e}\simeq 2.9 \times 10^{48}  \rm erg$, whereas
the energy in the electrons escaped the rim is 
$W_{\rm e}\simeq 1.9 \times 10^{48}  \rm erg$. It should be noted that the spectrum
of electrons outside of the rim is enriched  by multi-TeV particles 
due to the effect of energy dependent ($\propto E$) diffusive escape. 
The total energy in electrons required to explain the X-ray and 
TeV $\gamma$-ray fluxes is about $5 \times 10^{48}$ erg.  
Since the total energy of electrons depends on the
B-field as $W_{\rm e} \propto B^{-(1+\alpha_{\rm r})}$ the total electron energy could
be reduced only by a factor of $(8/6)^{1.6} \simeq 1.6$ for the magnetic 
field $B=8 \, \mu \rm G$. Therefore 
the inverse Compton origin of the detected 
TeV radiation requires about 1 percent of the total explosion energy of SN 1006
in relativistic electrons. 

The total energy of the magnetic field in the remnant with 
$B_2= B_1/4 \simeq  2 \, \mu \rm G$ for the distance 1.8 kpc is 
about $10^{46} \, \rm erg$.  A similar  amount of magnetic field 
energy is expected also in the rim since  the amplification of the
B-field there is compensated by a smaller volume of the rim.  
Thus,  the inverse Compton origin of TeV $\gamma$-radiation
implies that the conditions in SN 1006 are far from the equipartition
between relativistic electrons and the magnetic field.

In all calculations above we have assumed a stationary injection rate of
relativistic electrons which is a good approximation as far as 
the source is in an early Sedov phase. In order to demonstrate that, 
in Fig.~6 we show the spectra of synchrotron and IC fluxes calculated
for two different assumptions: (i) a time-independent injection rate 
$Q(t)=\rm const$, (ii) injection rate in the form $Q(t) \propto (1+t/t _\ast)^{-1}$
with a  characteristic time (onset of the Sedov phase) $t _\ast=100 \, \rm yr$.
In both cases we have assumed  acceleration spectrum given by Eq.~(1)
with  $E_0=30 \, \rm TeV$.  For the assumed  magnetic field 
in the rim   $B_1=7 \, \mu \rm G$  and the current shock speed
$v_{\rm s}=3600 \, \rm km/s$  this value of $E_0$ implies  a
gyrofactor $\eta=1.5$  (see Eq.~11).  
As it is seen from Fig.~6 the higher injection rate in the past does not affect
significantly the fluxes of nonthermal radiation even for a very small 
values of $t_\ast=100 \, \rm yr$. The main difference between the stationary 
and the time-dependent injection cases is reduced to somewhat  
higher IC $\gamma$-ray fluxes outside of the rim. For more realistic 
values of $t_\ast \geq  300 \, \rm yr$ the effect of time-dependent  injection 
becomes practically  negligible. 

%f6
\begin{figure}[htbp]
%\hspace*{7cm}
\resizebox{8.cm}{!}{\includegraphics{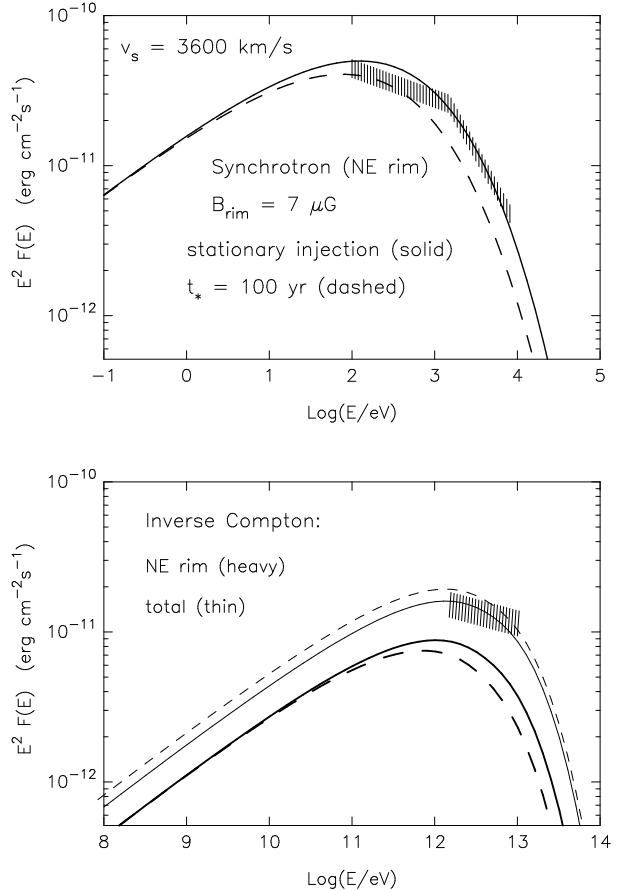}}
\caption
{The synchrotron (top panel) and IC  $\gamma$-ray (bottom panel)
fluxes calculated in the framework of 2-zone model for two 
assumptions about the injection rate of electrons:
$Q(t)=\rm const$ (solid lines), and  $Q(t) \propto (1+t/t _\ast)^{-1}$ with
$t_\ast=100 \, \rm yr$. In both cases $B=7 \, \mu \rm G$ 
and $E_0=30 \, \rm TeV$.
The distance to the source is 1.8 kpc.} 
\label{f6}
\end{figure}  

The above calculations show that the electronic origin 
of the observed TeV $\gamma$-ray emission meets  difficulties
for small distances to the source, $d \simeq  1 \, \rm kpc$ or less,
since the latter implies a small shock speed, and therefore insufficient
acceleration rate in order to accelerate  electrons  to  energies of about 
30 TeV and beyond.   In case of confirmation of the recent claims
about small distance to the source (Willingale et al. \cite{ROSAT}), 
a possible solution to this difficulty could be an 
assumption of a more effective 
acceleration mechanism compared with the
conventional diffusive shock acceleration,   
e.g. by invoking the mechanism of perpendicular 
diffusion proposed by Jokipii \cite{Jok}. 
Another principal possibility to increase the  accelerate rate  is connected 
with an assumption of a   large magnetic field in the rim, 
$B_1 \gg 10 \, \rm \mu G$. Obviously in this scenario
the TeV radiation principally cannot be explained
by the IC mechanism. The only reasonable alternative 
could be the  nucleonic origin of TeV radiation.   

\section{TeV gamma rays of nucleonic origin}

Currently,  the nucleonic origin of the observed TeV emission
is treated by the community as an inadequate  alternative
to the IC mechanism,
the main argument being the low ambient density of the 
gas in SN 1006, $n \simeq 0.4 \, \rm cm^{-3}$ (Willingale et al. \cite{ROSAT})
as well as the supposed large distance to the source of 
about 2 kpc  (Winkler and Long \cite{winkler}). 
Hower these arguments are not sufficiently robust in order to dismiss
such an important possibility with far going conclusions
concerning the origin of the nucleonic component 
of galactic cosmic rays (Aharonian \cite{vertalk}). 

\begin{figure}[htbp]
%\hspace*{7.cm}
\resizebox{8.cm}{!}{\includegraphics{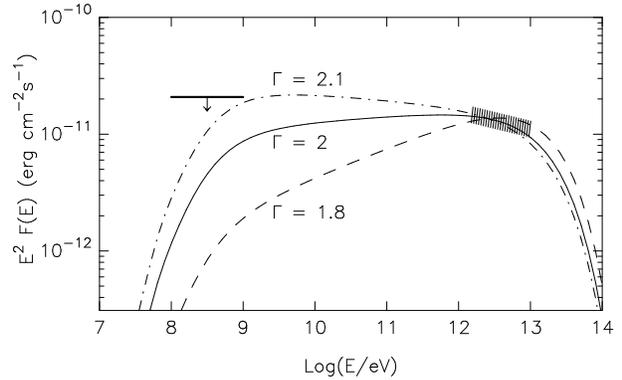}}
\caption
{The fluxes of $\pi^0$-decay $\gamma$-rays calculated for the
proton spectrum in  Eq.(1)  with $E =200  \, \rm TeV$,
and 3 different power-law indices: $\Gamma_{\rm p}=1.8$
(dashed), 2 (solid),  and 2.1 (dot-dashed).  The fluxes
are normalized to the observed flux  at
3 TeV. This results in the values of scaling parameter  $A=0.44$, 0.83, and 1.35,
respectively.}
\label{f7}
\end{figure}

Indeed, the very fact of the existence of 
$\gg 10 \, \rm TeV$  electrons as it follows from the X-ray data 
is an evidence for a strong shock in SN 1006, and  
implies a large compression  factor, $\rho \simeq 4$
or even more, up to 10,  as it follows from  
non-linear studies of shock acceleration in
SNRs (Malkov \cite{misha}, Berezhko \cite{berezh96}, 
Berezhko \& V\"olk \cite{BVLK}, V\"olk \cite{KPVLK}, 
Baring et al. \cite{baring}).  
Therefore it is not unrealistic if one
assumes a significantly higher density of the gas in the 
rim region, e.g.  $n \simeq 2 \, \rm cm^{-3}$. 
The current estimates of the distance to the source also contain 
large uncertainties, and do not exclude a distance of about 1 kpc or even less.
In particular, one of the recent studies 
of SN 1006 based on the ROSAT observations suggests  a  
distance $d=0.7 \pm 0.1 \, \rm kpc$
(Willingale et al. \cite{ROSAT}).
To explain  the observed flux of TeV emission,
for $\Gamma_{\rm p}=2$ the scaling factor  
$A \simeq 0.83$ is needed (Fig.~\ref{f7}). This requires  
for the total energy in accelerated protons 
\begin{equation}
W_{\rm p} \simeq 4  \times 10^{49} \,  (n/2 \,  
{\rm cm^{-3}})^{-1} \, d_{\rm kpc}^2 \, \rm erg \, .
\end{equation}
This is only $\sim 10$ per cent   of the  total kinetic energy 
of explosion estimated for SN 1006 as 
$5 \times 10^{50} \, \rm erg$.  The numerical calculations  
within the kinetic theory of  shock acceleration 
show  (Berezhko \cite{berezh99}) 
that even for a large 
distance to  SN~1006, the energy budget in
accelerated protons can be sufficient to explain the TeV $\gamma$-ray emission
from SN~1006 by hadronic interactions.

%f7

In fact this estimate of the scaling factor A derived 
from the comparison of calculated and observed TeV fluxes 
depends significantly on the  spectrum of protons. 
In Fig.~7 we show  the $\pi^0$-decay
$\gamma$-ray fluxes calculated for 3 spectra of 
protons taken in the form of Eq. (1)  with spectral
index $\Gamma_{\rm p}=$1.8, 2, and 2.1.
The fluxes are normalized to the observed flux  at 3 TeV.
The corresponding values of the scaling factor
are A=0.44, 0.83, and 1.35, respectively. 
Even in the case of relatively soft spectrum of protons 
with $\Gamma_{\rm p}=2.1$, the required scaling factor is 
still acceptable taking into account that 
more than  $20 \%$ of the energy of the supernova 
explosion can be transformed into  accelerated protons
(Berezhko \& V\"olk  \cite{BVLK}). Steeper spectra of protons
with a slope $\Gamma_{\rm p} \ge 2.1$ do not match  
the energy budget of the source.
Such spectra are excluded also by the EGRET flux upper 
limit, $F_\gamma \leq 1.7 \cdot 10^{-7} 
\, \rm ph/cm^2 s$ at $E \geq 100 \, \rm MeV$ (Mori \cite{Mori}).

%f8
\begin{figure}[htbp]
%\hspace*{7.cm}
\resizebox{8.cm}{!}{\includegraphics{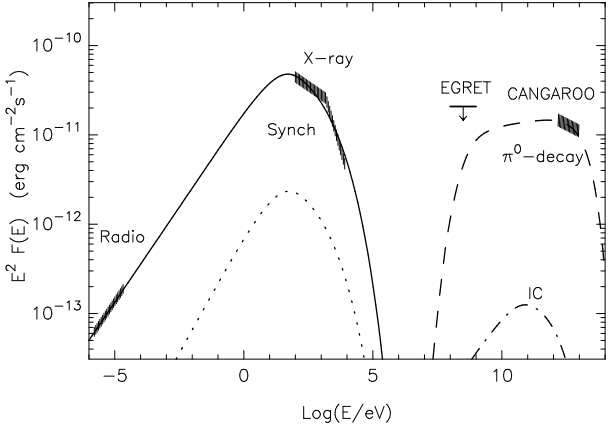}}
\caption
{The nonthermal radiation of SN 1006 
calculated for large magnetic field in the NE rim $B=70 \, \rm \mu G$.
Solid curves correspond to the synchrotron radiation of directly 
accelerated electrons, while the dotted curve 
corresponds to the synchrotron radiation produced  by secondary
($\pi^{\pm}$-decay) electrons.   The dashed curve shows the 
flux of $\pi^0$-decay $\gamma$-rays calculated for
proton spectrum with $\Gamma_{\rm p}=2$, $E_0=200 \, \rm TeV$,
and  scaling parameter $A=0.83$.  The dot-dashed line shows the
flux of IC $\gamma$-rays contributed mostly by directly
accelerated electrons.  All fluxes  shown are well dominated by the 
emission produced in the rim.} 
\label{f8}
\end{figure}

The $\gamma$-ray spectra shown in Fig.~7 are calculated
assuming  for the maximum 
energy of protons $E_0=200 \, \rm TeV$.
Although the exact value of $E_0 $
do not noticeably change the requirements to the 
scaling factor $A$,   it has a noticeable impact on the
spectral form of $\gamma$-rays above 10 TeV. 
Therefore only future precise spectroscopic measurements 
in this energy region could provide an important information
about $E_0$. At the same time the existing sub-10 TeV data  
already tell us that within the framework of nucleonic 
model of TeV radiation of SN 1006 the cutoff energy 
$E_0$ could not be significantly less than 100 TeV.   

Although in this paper we do not attempt to 
discuss the aspects related to the theory of
diffusive shock acceleration in SNRs, it 
is clear that such large values of $E_0$ could be achieved 
only under the assumption of a large magnetic field in the
shock region, $B \sim 10^{-4} \, \rm G$ or even more
(see Eq.~11). Obviously this assumption does not leave any
chance for the IC mechanism to give  a noticeable 
contribution into the observed TeV emission. 
On the other hand the featureless X-ray radiation could
be naturally explained by the synchrotron radiation
of accelerated electrons. In such strong magnetic fields
the maximum energy of accelerated electrons is determined
by the balance between the acceleration rate and the
synchrotron energy loss rate:
\begin{equation}
E_0 \simeq 45  \, \eta^{-1/2}    \left(\frac{
B}{10 \, \rm {\mu G}}\right)^{-1/2} 
\left(\frac{v_{\rm s}}{2000 \, \rm km/s}\right )
\; \rm TeV \, ,
\end{equation}   
and therefore,  as it follows  from Eq.(8)  the energy of exponential cutoff in the 
synchrotron spectrum,
\begin{equation}
E_{\rm m}  \simeq 0.5 (v_{\rm s}/2000 \, \rm km/s)^2 
\eta^{-1} \, \rm  keV \, .
\end{equation}
does not depend on neither the magnetic field nor the age of
a SNR. The very fact of detection of nonthermal X-rays 
above 1 keV testifies that the particle acceleration 
takes place  in the limit of Bohm diffusion ($\eta \sim 1$).  

Therefore an observation of flat X-ray spectrum
in the region 1-10 keV from any SNR would imply, for 
a reasonable  shock speed in the Sedov phase 
$v_{\rm s} \leq 4000 \, \rm km/s$,  a more effective
acceleration than it follows from the 
prediction of conventional parallel shock acceleration model.

In the regime of non-negligible synchrotron losses 
we should expect, in addition to the exponential cutoff at
$E_{\rm m}$, another spectral feature -- a spectral 
break (increase of the power-law index by 0.5)
at the  energy $E_{\rm b}$ which is determined from the 
condition $t_{\rm sy}=t_0$:
\begin{equation}
E_{\rm b} \simeq 2.9  (B/10 \, {\rm \mu G})^{-3}
\, (t_0/10^3 \, {\rm yr})^{-2} \, \rm keV. 
\end{equation}

In Fig.~\ref{f8}  we show the spectrum of synchrotron radiation calculated for
the following parameters: $B = 70 \, \rm \mu G$,
$t_0 = 700 \, \rm yr$, $v_{\rm s} = 2900 \, \rm km/s$,
and $\eta=1$. For these parameters the spectral break
is  at $E_{\rm b} \simeq 0.02 \, \rm keV$, while 
the exponential cutoff $E_{\rm m} \simeq 1 \, \rm keV$.
Although there is a gap by a factor of 20 between  these two
characteristic photon energies, the additional gradual steepening
due to the exponential cutoff in the electron spectrum
becomes noticeable already at energies $E_{\rm x} \geq 0.1 E_{\rm m}$.

\section{Conclusions}

In the previous theoretical studies of the IC $\gamma$-radiation of SN 1006
a  homogeneous one-zone model  has been implied (Mastichiadis \& de Jager  \cite{masti},
Pohl \cite{pohl},  Yoshida \& Yanagita \cite{yoshida}) 
which assumes that  all accelerated electrons are confined in the NE rim.  
The magnetic field in the rim derived from these studies which would  not contradict
the observed X-ray to TeV $\gamma$-ray flux ratio should not exceed 
$7 \, \rm \mu G$.    

Our study of the  $\gamma$-ray production in SN 1006
which includes the effect of convective and diffusive escape of 
electrons,  allows  some increase of the magnetic field 
up to  10 $\rm \mu G$, with preferable values being in a 
rather narrow   range between 6 and 8 $\rm \mu G$. 
Given the short time of particle acceleration 
available, $t_0 \le t_{\rm SN 1006} \simeq 10^3 \,\rm yr$, 
the assumption of such a small magnetic field
requires a very  high shock speed in SN 1006, $v_{\rm s} \geq 3000 \, \rm km/s$,
in order to provide acceleration of electrons to energies $E_0\sim  30 \,\rm TeV$
needed for explanation of X-ray and TeV observations. This implies a {\it large}
distance to the source, $d \geq 1.5 \, \rm kpc$.   

This model allows definite predictions which could be tested by future observations.
A significant, if not dominant, fraction of the IC TeV emission is to be produced
outside of the NE rim.  Namely,  a $\gamma$-ray flux comparable with the flux from
NE rim, should be then expected from the extended inner region  of the remnant 
adjacent to the rim. The size  of this region depends on the character of propagation
of electrons in the rim.  Indeed, even in the case of 
diffusion of electrons in the Bohm limit in the remnant with magnetic field  
$B_2 \sim 2 \, \rm \mu \rm G$,  the characteristic distance of penetration of electrons 
towards the central region of the remnant 
could be estimated as 
\begin{equation}
l(E_{\rm e}) \simeq \sqrt{2 \, D(E_{\rm e}) t_0} \simeq 0.75 \, 
(E_{\rm e}/10 \, \rm TeV)^{1/2} \, \rm pc .
\end{equation}

Since the  IC production of TeV $\gamma$-rays (on 2.7 K MBR) 
with energies  less than several TeV takes place in the Thompson regime, 
and therefore their characteristic energy  
scales as $E \simeq 1 \, (E_{\rm e}/20 \, \rm TeV)^2 \, \rm TeV$, 
the size of the emission region of TeV gamma rays should be as large as the
width  of the rim, and  increases  linearly with energy of $\gamma$-rays.
Moreover,  if the propagation of electrons in the remnant is  much faster
than in the Bohm limit ($\eta \gg 1$),  the electrons could fill up practically 
the entire  remnant.  In that case the energy dependence of the 
size of the $\gamma$-ray emission will be significantly weakened. 

Another distinctive feature of the IC origin of TeV radiation is the spectral
shape of the radiation: very hard, with the photon index 
$\alpha_\gamma\sim 1.5$ below 1 TeV, a flat spectrum with 
$\alpha_\gamma\sim 2$ between 1 and 10 TeV, and very steep above  10 TeV. 

The alternative to the IC interpretation  is the nucleonic origin of the observed 
TeV radiation. This interpretation becomes energetically comfortable,  which
implies no more than $10^{50}\,\rm erg$ in accelerated protons and nuclei,
if we assume (i) a {\it small} distance to the source of about 1 kpc or less,
and (ii)  a significant enhancement of the gas density in the rim due 
to strong shock compression. This predicts a compact size of the TeV emission
comparable with that of the X-ray rim. In the region below 1 TeV  
the spectrum of $\gamma$-rays of nucleonic origin,  with a power-law 
index $\alpha_{\gamma}\simeq \Gamma\sim 2$  is expected 
steeper than the spectrum of IC $\gamma$-rays. 
However  a flatter spectrum of $\pi^0$-decay  $\gamma$-rays 
cannot be excluded if the protons would  have an acceleration spectrum
as hard as  $E^{-1.5}$ (see Malkov \cite{misha}). 

The shape of the
spectrum of $\pi^0$-decay $\gamma$-rays above 1 TeV depends 
on the characteristic  maximum energy of accelerated protons. 
Since the flux of TeV $\gamma$-rays
is no more connected with  the X-ray fluxes, 
the value of the magnetic field in the acceleration region (NE rim) could be adopted 
as high as $\sim 100\,\rm \mu G$. Correspondingly, the energies 
$E_0\geq 100\,\rm TeV $ could be achieved. On the other hand if $E_0$
would be significantly less than 100 TeV, one may expect a
turnover in the spectrum of $\gamma$-rays above several TeV.   

Thus, it is quite possible that both in the low 
(sub-TeV)  and in the high (multi-TeV) energy regions  
the IC and $\pi^0$-decay $\gamma$-rays may
have similar spectra. 
Therefore the spatial rather than spectroscopic measurements 
of $\gamma$-radiation above 100 GeV with future IACT arrays could provide 
decisive information about the origin  of very high energy radiation of  SN 1006.    
The IC models predict significant $\gamma$-ray fluxes not only
from the rim, but also from the inner parts of the remnant.
Meanwhile, the $\pi^0$-decay $\gamma$-rays trace the density 
profile of the gas in the production region,  therefore one should expect
a rather compact $\gamma$-ray source 
essentially coinciding with the rim. 

We emphasize that both the electronic and the nucleonic models of 
the observed TeV radiation of SN 1006 require a very effective acceleration
of particles in the regime close to the Bohm limit. Also,  the both models 
predict a very high conversion efficiency of the total kinetic energy of
the explosion available ($\sim 5 \times 10^{50} \, \rm erg$)
into relativistic particles. The energy required for relativistic electrons  
within  the framework of the IC model is predicted with a
good accuracy,  given the small range of allowed B-field:
$W_{\rm e} \simeq 5 \times 10^{48} \, \rm erg$.  This is by two orders of magnitude 
larger than the energy contained in the magnetic field;
this suggests that the  conditions in SN 1006  are far from the equipartition regime.

The nucleonic model of TeV emission requires total energy in accelerated 
protons from  $2 \times 10^{50} \, \rm erg$ down to  $10^{49} \, \rm erg$,
depending on the enhancement factor for the gas density in the rim, 
the spectral index of the accelerated protons, and the distance to the source. 
In particular for moderate assumption for the spectral index 
$\Gamma_{\rm p}=2$,  and the gas compression ratio $\rho=4$ 
(i.e. $n=1.6 \, \rm cm^{-3}$),  the energy in  protons
is estimated $\approx 5 \times 10^{49} (d/1 \, \rm kpc)^2 \, \rm erg$. 
Note that since the magnetic field in the rim for the nucleonic  model could be 
as high as $100 \, \mu \rm G$, this model predicts a situation
close to the equipartition.  

High magnetic fields in the rim can equally well explain the current X-ray data.
The synchrotron X-radiation in this case is produced in the regime of saturated
energy losses of electrons, which results in a somewhat different spectral shape
of the synchrotron fluxes. In both large and small  magnetic field limits,
the spectrum would be best fitted in the form $F(E)\propto E^{-\alpha_{\rm x}}
\exp[-(E/E_\ast)^{-1/2}]$, where $\alpha_{\rm x}\simeq 2.1$ 
and $\alpha_{\rm x}\simeq 1.6$ for  the cases of large and small magnetic fields,
respectively.

\begin{acknowledgements}  We express our thanks to  H. J. V\"{o}lk, M. Malkov,  E. Berezhko, 
T. Kifune and T. Tanimori  for very fruitful discussions. The work of AMA was 
supported through the Verbundforschung 
Astronomie/Astrophysik of the German BMBF under grant No. 05-2HD66A(7). 
\end{acknowledgements}

{}
\end{document}